\documentclass[12pt]{iopart}

\usepackage{graphicx,psfrag,bm}
\newcommand{\subfig}[3]{\raisebox{#1}{\hspace{-#2}\hspace{-1em}(#3)\hspace{#2}}}
\newcommand{\eqref}[1]{(\ref{#1})}
\newcommand{\lapprox}{%
\mathrel{%
\setbox0=\hbox{$<$}
\raise0.35ex\copy0\kern-\wd0
\lower0.65ex\hbox{$\sim$}
}}
\newcommand{\be}{\begin{equation}}
\newcommand{\ee}{\end{equation}}

\newcommand{\pder}[3]{\frac{\partial^{#1}#2}{\partial #3^{#1}}}
\newcommand{\beshpm}[2]{\mbox{H}_{#1}^{\pm}(#2)}
\newcommand{\besj}[2]{\mbox{J}_{#1}(#2)}
\newcommand{\beshp}[2]{\mbox{H}_{#1}^{+}(#2)}

\newcommand{\ex}[1]{\mbox{e}^{#1}}
\newcommand{\eps}{\epsilon}
\newcommand{\refcite}{\cite}
\newcommand{\braket}[2]{\langle #1 | #2\rangle}
\newcommand{\mcal}[1]{\mathcal{#1}}
\def\bal#1\eal{\begin{align}#1\end{align}}

\begin{document}
\title[An efficient Fredholm method for calculation of highly excited states of billiards]{An efficient Fredholm method for calculation of highly excited states of billiards}
\author{Hakan E.\ T{\"u}reci$^1$ and Harald G.\ L.\ Schwefel$^2$}
\address{$^1$ Institute of Quantum Electronics, ETH Z\"urich, 8093 Z\"urich, Switzerland}
\address{$^2$ Max-Planck Research Group for Optics, Information and Photonics, Univ.\ of Erlangen, G.-Scharowsky-Str. 1, 91058 Erlangen, Germany}
\ead{hschwefel@optik.uni-erlangen.de}

\begin{abstract} 
A numerically efficient Fredholm formulation of the billiard problem is presented. The standard solution in the framework of the boundary integral method in terms of a search for roots of a secular determinant is reviewed first. We next reformulate the singularity condition in terms of a flow in the space of an auxiliary one-parameter family of eigenproblems and argue that the eigenvalues and eigenfunctions are analytic functions within a certain domain. Based on this analytic behavior we present a numerical algorithm to compute a range of billiard eigenvalues and associated eigenvectors by only two diagonalizations.
\end{abstract}

\submitto{\JPA}
\maketitle

\section{Introduction}
 
The billiard problem has played a vital role in the study of the manifestations of classical chaos in linear wave systems (``wave chaos") including microwave, optical and acoustic cavities  and waveguides~\cite{stockmann1992, blumel1992, soap98, science98, Doya2001, Reichl2003, kuhl2005, wiersig2006}, and various single-particle quantum systems~\cite{stockmann_book, haake_book, Y2Kproc_book}. Even in strongly-interacting, non-linear systems the knowledge of the linear spectrum and eigenfunctions is paramount to infer complex observables \cite{Kurland2000,AgamA01,Stafford2003,Zelevinsky2006, harayama2005, tureci2006}. In the semiclassical limit, or at high wavenumbers ($k=2\pi/\lambda$), the numerical solution of the Laplace eigenvalue problem becomes computationally challenging. Finite difference schemes~\cite{Morton2005} become impractical and Green's function matching methods~\cite{alex_thesis} suffer from the unfeasibility of a root search.

The typical Green's function matching method (various implementations of which includes the Method of Particular Solutions (MPS) and boundary integral methods (BIM)) to solve the Laplace eigenproblem consists of finding the zeros of the secular determinant over a given wavenumber range. In practice, this is accomplished through the singular value decomposition (SVD) and scanning for the minima of the smallest singular values~\cite{backer2003}. This requires typically of the order of $(kR)^3$ matrix operations per mode (where $R$ is the typical size of the system). Naturally, this procedure becomes progressively more expensive for higher lying eigenvalues. Missing eigenvalues are a more important problem in practice. At larger wavenumbers, when the spectrum becomes progressively denser, it's a serious problem to differentiate and separate the minima of the lowest singular values.\footnote[7]{Nearly degenerate levels can in practice be differentiated within the SVD scheme by looking at several of the smallest singular values~\cite{backer2003}.} 

In this paper, we propose a fast and efficient method based on a Fredholm formulation of the billiard problem, to compute the spectrum and the corresponding eigenfunctions of the Laplace operator over a two-dimensional domain $D$. This method is closely related to the scattering quantization method (SQM)~\cite{DietzEPSU95,Frischat97,Tureci05} as it relies on a similar acceleration technique of replacing the search for singular values of a matrix by an auxiliary eigenvalue problem. In contrast to SQM which expands the Laplace eigenfunctions in terms of a set of basis functions of the Laplace operator in the domain $D$, the expansion here contains the fundamental solutions of the Laplace operator. This has two important advantages which makes its exposition worthwhile. First, the proposed Fredholm formulation is known to be uniformly convergent~\cite{atkinson97numerical} while the SQM is known to be convergent only in so far as the Rayleigh hypothesis holds~\cite{tureci_thesis}. Second, Fredholm formulations via BIM are amenable to semiclassical quantization techniques through the transfer operator technique. Consequently, the behavior of the Laplace operator for various domain geometries in the semiclassical limit can be directly related to the invariants of underlying classical motion in that domain~\cite{TasakiHS97, Simonotti2000, Prange1995}.

We would like to remark that the method outlined here provides a similar gain in speed and robustness with respect to the scaling method of Vergini and Saraceno~\cite{VerginiS95,Barnett2006}. A recent boundary integral formulation of the scaling method has been carried out in~\cite{Robnik2007}. To the authors' best knowledge the relation between scattering quantization methods and scaling methods is still an open question. 

\section{Review of the BIM formulation}

Let us briefly review the BIM formulation of the billiard problem that we are addressing. Consider a two-dimensional Euclidean domain $D$ bounded by a smooth boundary curve $\partial D$. Within this domain, let $\{\psi_\mu(r)\}$ be the set of eigenfunctions of the Laplace operator with eigenvalues $k_\mu^2$,
\be
-\nabla^2 \psi_\mu(\bm{r}) = k_\mu^2 \psi_\mu(\bm{r}) .
\label{eqlaplaceeig1}
\ee
We assume that $\psi_\mu(r)$ satisfies Dirichlet boundary conditions $\psi_\mu|_{\partial D}=0$. In the context of the Schr\"odinger equation, $E_\mu =k_\mu^2$ are the discrete energy levels of a particle in a box defined by $\partial D$.

Associated to the differential operator in the above equation~\eqref{eqlaplaceeig1} is the Green's function 
\be
(\nabla^2 + k^2) G(\bm{r},\bm{r}';k) = \delta (\bm{r}-\bm{r}').
\label{eqgreen0}
\ee
Regardless of the boundary conditions on the Green's function one can reformulate the billiard problem~\eqref{eqlaplaceeig1}, through a completely standard procedure, in terms of a Fredholm integral equation of the second kind
\be
\int_{\partial D} d\bm{r}(s') \pder{}{}{n}G (\bm{r}(s),\bm{r}(s');k) \pder{}{}{n'}\psi(\bm{r}(s')) = -\frac{1}{2}\pder{}{}{n}\psi(\bm{r}(s)),
\label{eqdirichbim1}
\ee
which has solutions only for discrete values $k=k_\mu$. In the above equation, $s$ is the arc length along the boundary, $\partial / \partial n = \bm{n}(s) \cdot \bm{\nabla}$, $\partial / \partial n' = \bm{n}(s') \cdot \bm{\nabla}$, and $\bm{n}(s)$ is the outwards pointing unit normal of the boundary at location $s$.

\begin{figure}[h!]
\centering
\psfrag{n}{$\bm{n}(s)$}\psfrag{n'}{$\bm{n}(s')$}
\psfrag{s'}{$\bm{r}(s')$}\psfrag{s}{$\bm{r}(s)$}
\psfrag{T}{$\theta(s,s')$}\psfrag{T'}{$\theta(s',s)$}
\psfrag{l}{$l(s,s')$}
\includegraphics[clip,width=0.45\linewidth]{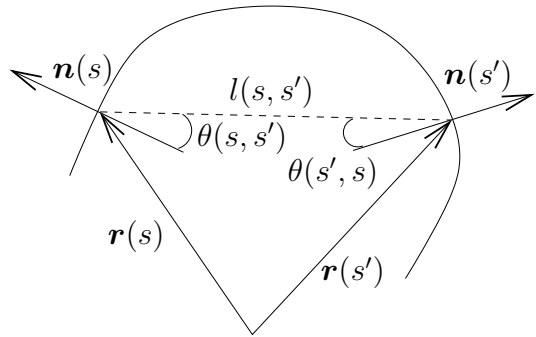}
\caption{Schematics showing the variables used in the definition of the BIM kernel in equation~\eqref{eqBIMkerneldef}.}
\label{fig:schem}
\end{figure}

Therefore, the problem in the two-dimensional domain is reduced to a problem on the boundary. This reduction is physically very appealing as in the semiclassical limit the geodesic flow is uniquely represented as a discrete map on the boundary. Of course, the reduction in dimensionality has certain consequences. Whereas in the standard treatment of domain problems through finite element methods one solves for the whole spectrum up to a maximal wavenumber $k$, boundary formulations provide a narrow spectral range around a reference wavenumber $k$.

The standard BIM formulation employs the free-space outgoing Green's function~\cite{ backer2003,Boasman94,KosztinS97}
\be
G_0(\bm{r},\bm{r}'; k) = - \frac{i}{4} \beshpm{0}{k|\bm{r}-\bm{r}'|}.
\label{eqgreenfn0}
\ee
Here $\beshpm{0}{z}$ are the first and second kind Hankel functions of order zero. Let us rewrite the Fredholm problem~\eqref{eqdirichbim1} in an operator notation 
\begin{equation}
K u = u,
\end{equation}
where $u(s) = \pder{}{}{n}\psi(\bm{r}(s))$ and the kernel, using the free Green's function~\eqref{eqgreenfn0} becomes
\begin{eqnarray}
K(s,s';k) &= -2\pder{}{G_0(\bm{r}(s),\bm{r}(s');k)}{n}\nonumber \\
&= -\frac{ik}{2} \cos\theta(s,s') \beshp{1}{k|\bm{r}(s)-\bm{r}(s')|}.
\label{eqBIMkerneldef}
\end{eqnarray}
Here, $\cos\theta(s,s') = \bm{n}(s)\cdot (\bm{r}(s)-\bm{r}(s'))/|\bm{r}(s)-\bm{r}(s')|$, i.e.\ $\theta(s,s')$ is the angle between the normal at $s$ and the cord connecting $s$ and $s'$ (see figure~\ref{fig:schem}). Consequently, $K$ (referred to as $K(k)$ in alternative notation) is clearly not a symmetric operator. Note that the diagonal elements are finite and given by
\be
\lim_{s\rightarrow s'} K(s,s';k) = \frac{1}{2\pi}\kappa(s),
\label{eqKerneldiag}
\ee
where $\kappa(s)$ is the curvature at $s$. Hence the condition of quantization is 
\be
\det(1-K(k)) = 0.
\label{eqsecdet1}
\ee
The standard numerical procedure to extract the zeros of this secular determinant in the context of billiards is outlined in~\refcite{backer2003,KosztinS97}. 

\section{Scattering quantization approach to BIM}
\begin{figure}[h!]
\centering
\psfrag{number}{$\!\!\#/N$}
\psfrag{lambda}{$|\lambda|$}
\psfrag{Im}{Im$[\lambda]$}
\psfrag{Re}{Re$[\lambda]$}
\includegraphics[clip,width=0.5\linewidth]{./figure2a.eps}\subfig{-1ex}{0.5\linewidth}{a}\\
\includegraphics[clip,width=0.5\linewidth]{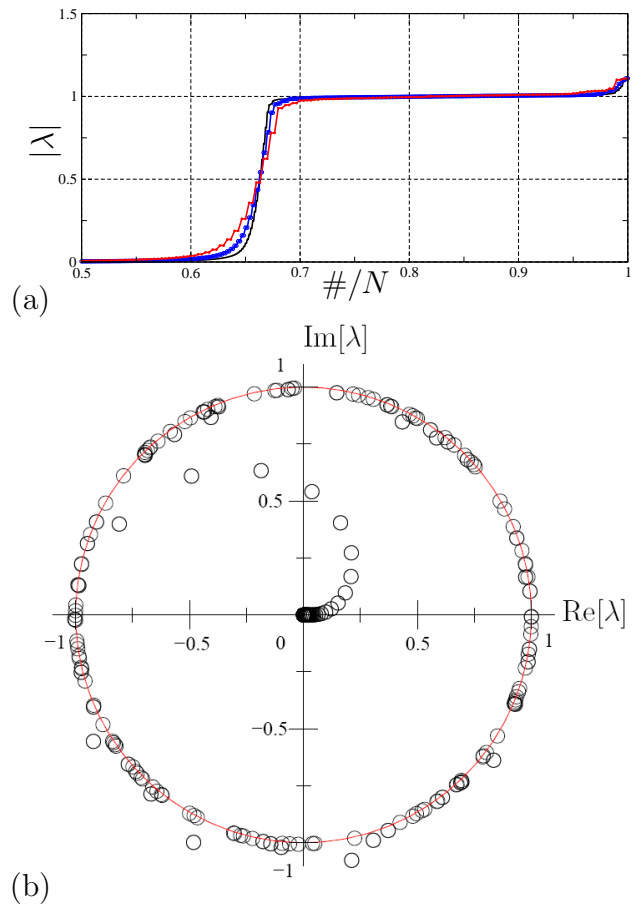}\subfig{-1ex}{0.5\linewidth}{b}
\caption{(a) Distribution of the magnitude of the eigenvalues of $K(k)$ for $kR=20$ (red), $100$ (blue), $200$ (black) for a quadrupolar billiard ($R(\phi)=R(1+\eps\cos2\phi)$) of deformation $\eps=0.1$. The eigenvalues are ordered with respect to their absolute value, and the horizontal axis denotes their relative order within all the eigenvalues ($N$). Note that the unitary sector scales linearly with $k$ (corresponds approximately to the number of open classical channels which can be estimated to be $2[kR]$). As the number of eigenvalues scales with the size of the system too, the unitarity border is identical in all cases. (b) Distribution of the eigenvalues in the complex plane for $kR=100$. The solid line is the unit circle. In each case, the size of the system and hence the number of eigenvalues is $N=[6\times kR]$.}
\label{figsqevals1}
\end{figure}

In contrast to the standard procedure outlined in the last section, we shall reformulate the problem by considering the solution of the eigenvalue problem
\be
K(k)u = \lambda u.
\label{eqeigKsimp1}
\ee
This eigenvalue problem provides us with a set of eigenvalues and eigenfunctions, $\{\lambda^{(i)}(k),  u^{(i)}(k;s)\}$ parametrically dependent on the continuous variable $k$. The structure of the operator $K(k)$ is interesting. It can be shown via stationary phase integration that in the semiclassical limit, $KK^{\dagger}$ is asymptotically diagonal i.e.\ while the off-diagonal elements are $O(\sqrt{k})$, the diagonal elements are $O(k)$. The form of the diagonal elements is given by 
\begin{equation}
(KK^{\dagger})_{ss} =  \frac{k}{2i \pi} \, \int ds' \,  \frac{\cos^2\theta(s,s')}{l(s,s')}.
\end{equation}
For arbitrary shapes, $K(s,s';k)$ is however not unitary \cite{bogomolny92} and does not obey the spectral theorem. 

Nevertheless, a favorable property of this set is that for the finite-dimensional truncation of $K(k)$, the spectrum can be roughly divided into a {\em null space} and a {\em unitary sector} (to be defined below). This can best be visualized by looking at the eigenvalue distribution of $K$. In figure~\ref{figsqevals1}(a), we plot the absolute values of the eigenvalues $\{\lambda^{(i)}(k_0)\}$. It's clearly seen that the distinction between null-space eigenvalues ($|\lambda^{(i)}| \approx 0$) and the unitary eigenvalues  ($|\lambda^{(i)}| \approx 1$) becomes sharper for larger $k$ i.e. in the semiclassical limit. At a typical value of $k$, the eigenvalues are distributed in the complex plane {\em within} the unit circle, and a fraction of the eigenvalues lie in the vicinity of the unit circle representing the unitary sector (see figure~\ref{figsqevals1}(b)). The size of this unitary sector is approximately $2[kR]$, which corresponds approximately to the number of half wavelengths on the boundary ~\cite{Doron92}. 

Returning to the eigenvalue equation~\eqref{eqeigKsimp1}, we see that the quantization condition~\eqref{eqsecdet1} can be rewritten as $\lambda(k_q)=1$. In other words, whenever we find an eigenvalue $\lambda^{(i)}(k_q)$ at $1+0i$ in the complex plane, $k_q$ is a solution of~\eqref{eqsecdet1} and $u^{(i)}(k_q;s)$ is the associated quantized eigenvector. 

\begin{figure}[t]
\centering
\psfrag{kR}{$kR$}
\psfrag{overlap}{\vspace{-10mm}$\!\!\!\braket{u^{(i)}(k_0)}{u^{(j)}(k)}$}
\psfrag{speed}{$d\phi/dkR$}
\includegraphics[clip,width=0.5\linewidth]{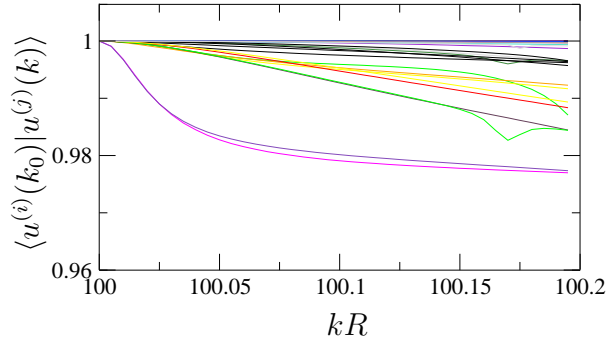}
\caption{Eigenvectors of the quadrupole with $\eps=0.1$. Overlap of the eigenvector $u^{(i)}(k_0)$ at $kR=100$ with the traced eigenvector $u^{(i)}(k)$ over the range $kR=[100.0,100.2]$.}
\label{fig:traceOverlap}
\end{figure}

We will now argue that not only are the (unitary sector) eigenvectors of $K(k)$ approximately orthogonal at a given $k$, but they also approximately diagonalize $K(k)$ over a range $\delta kR \sim O(1)$. (This range roughly corresponds to the one dimensional free spectral range of the billiard, which is roughly the number of wavelengths one can fit into the longest chord in the domain $D$. We will refer to it shortly as the ``free spectral range".) Consider the eigenvectors calculated at two different but close values of the parameter $k$, say $k_0$ and $k_0+\delta k$. We can define the overlap between the eigenvectors calculated at these two different values by  
\be  
\braket{u^{(i)}(k_0)}{u^{(j)}(k_0 + \delta k)} = \sum_{l=1}^{N} (u^{(i)})^*(k_0;s_l) \, u^{(j)}(k_0+\delta k;s_l).
\ee 
This operation is well-defined as long as we keep the system size $N$ constant. In figure~\ref{fig:traceOverlap}, we start with an initial set of states $|u^{(i)}(k_0)\rangle$, $i=1,\ldots,N$ and plot for subsequent $k=[k_0,k_0+\Delta k]$ only the overlap of the various initial states with their maximal overlap partner. We would like to note that there is in general only one state at $k$ that has a considerably larger overlap than all other states with an initial state $u^{(i)}(k_0)$. Here we plot only a fraction of the initial eigenvectors for the sake of visibility, but this behavior holds in general over stretches $\delta kR \sim O(1)$ of the parameter $k$ for eigenvalues in the first and fourth quadrant of the complex plane ($|\arg{\lambda}^{(i)}|<\pi/2$). The typical change in overlap over $\delta kR=0.2$ at $kR \simeq 100$ is less than $\%1$. 
\begin{figure}[t!]
\flushright
\psfrag{Im}{Im$[\lambda]$}
\psfrag{Re}{Re$[\lambda]$}
\psfrag{Error}{\hspace{-1mm}\raisebox{1ex}{$B_\mu$}}
\includegraphics[clip,width=12.0cm]{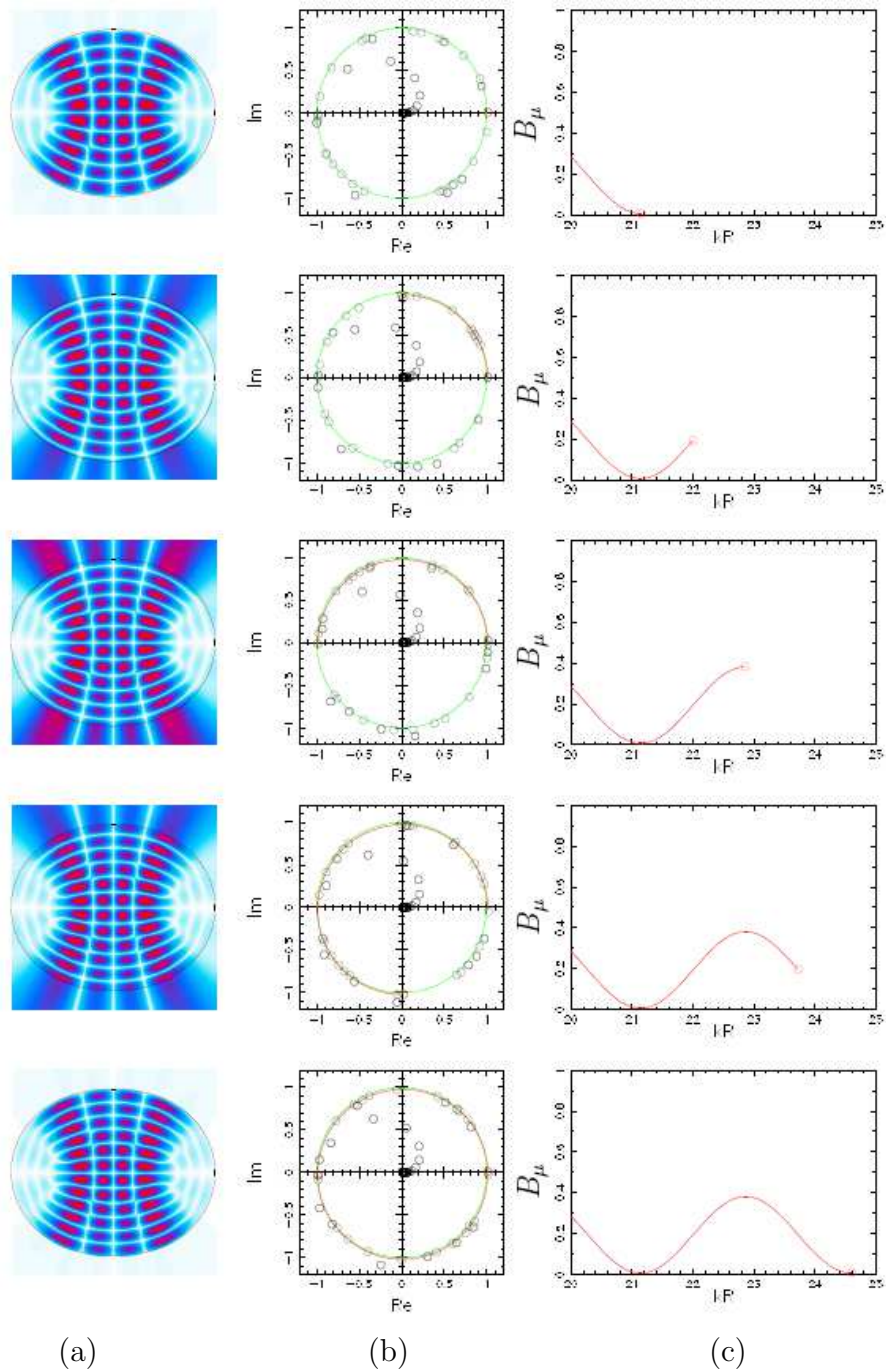}
\subfig{-3ex}{10.9cm}{a}\subfig{-3ex}{6.9cm}{b}\subfig{-3ex}{2.5cm}{c}
\caption{We follow an eigenfunction of the quadrupolar shape with $\eps=0.1$ via the method of highest eigenfunction overlap (see figure~\protect\ref{fig:traceOverlap}). Starting from the quantized eigenvalue at $kR=20.725$ we follow it in steps of $\delta kR=0.025$ through one quantization cycle up to $kR=24.175$. (a) False color plots of the intensity of the traced wavefunction. (b) Eigenvalues at each snapshot. In red we trace the motion of the eigenvalue of the particular state plotted in (a) and in (c) we plot the corresponding error on the boundary given by $B_\mu = \frac{1}{\mcal{L}} \oint_{\partial D} ds |\psi_\mu(\bm{r}(s))|^2$.}
\label{fig:trace1state}
\end{figure}

An important consequence of this observation is that we can assign an identity to the eigenvectors {\em even away from quantization}~\cite{Tureci05,Frischat95}. To elucidate this point, consider the trace of one of the eigenvectors in figure~\ref{fig:trace1state}. The initial eigenvector is not quantized and we follow this state by the highest-overlap criterion over a range of $\delta kR \approx 3.5$, a range that is larger than the free spectral range. We only plot here five instances over which the state becomes quantized (top to bottom). An important feature of this behavior is the way an extra node is ``pushed" into the billiard. We have to note that over such large stretches of $kR$, an eigenvector typically  undergoes avoided crossings. The avoided crossings happen predominantly around $\arg{\lambda}^{(i)} \approx \pm\pi$ in the second and third quadrants of the complex $\lambda$-plane. This is the region of the complex eigenvalue plane where the null-space eigenvalues join the ``unitary flow" (see figure~\ref{figsqevals1}(b) and the animation in the media section). However, the numerical method that we propose below utilizes the behavior in the first and fourth quadrants in the complex eigenvalue plane away from avoided crossings.

\begin{figure}[t]
\centering
\psfrag{kR}{$kR$}
\psfrag{overlap}{\hspace{-2ex}$\braket{d\psi_o}{d\psi_k}$}
\psfrag{speed}{$d\phi/dkR$}
\psfrag{dspeed}{max($d\phi/dkR$)/$d\phi/dkR$}
\includegraphics[clip,width=0.65\linewidth]{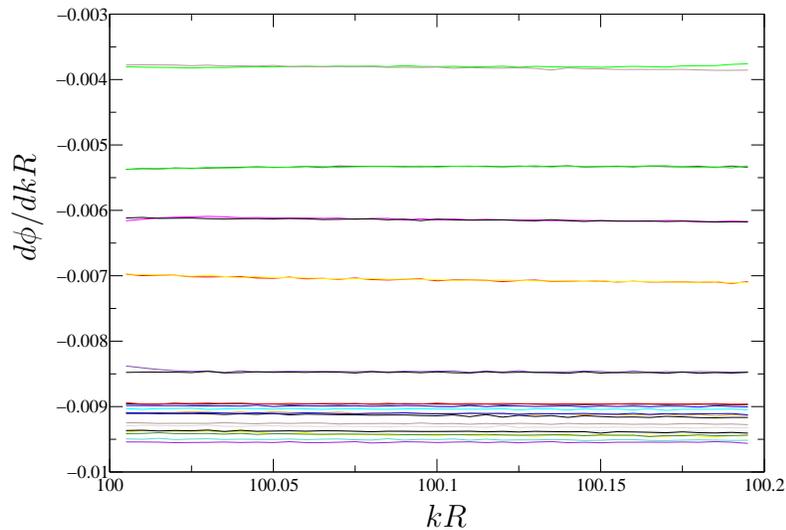}
\caption{Eigenvalues of the quadrupole with $\eps=0.1$. The speed $d\phi/dkR$ of a number of  eigenvalues around $kR=100$. The corresponding eigenvectors are traced over one fourth spectral range, $kR=[100.0,100.2]$.}
\label{fig:traceOverlapSpeed}\label{fig:evalspeeddist1}
\end{figure}

A second key observation concerns the behavior of eigenvalues $\lambda_\mu(k)$ of $K(k)$. This notation makes explicit the adiabatic identity of the eigenvectors that we have established above. With increasing $k$, the eigenvalue flow is counterclockwise. There is a clear distinction between the unitary eigenvalues which flow along the unit circle $|\lambda| = 1$ and the null-space vectors which accumulate at $\lambda \sim 0$. The eigenvalues in transition that have an intermediate value of $|\lambda|$ follow a universal path (compare to the case of circular billiard in figure~\ref{fig:analyticEigen}) and are added to the unitary flow at about $\phi \equiv \arg{\lambda} \approx \pi$ as noted above. This is the mechanism by which the density of states of the billiard eigenvalues increase, which according to the Weyl formula has the mean asymptotic behavior $\rho_{\mbox{\tiny\em Weyl}}(k) = k\mcal{A}/2\pi$, where $\mcal{A}$ is the area of the domain $D$. In figure~\ref{fig:traceOverlapSpeed} (a), we show that the phase speeds of the unitary eigenvectors, defined by $v^{\mu}_\phi(k) = d\phi_\mu(k)/dk$ is {\em constant} over a stretch of $\delta kR \sim O(1)$. This is one of the main ingredients of the numerical diagonalization procedure that we propose in the next section. 

\section{An accelerated Fredholm root search and the accuracy of solutions}

\begin{figure}[t]
\centering
\psfrag{error}{\raisebox{1ex}{$\!\!\!\!\!\!E_\mu(k_0)$}}
\includegraphics[clip,width=7cm]{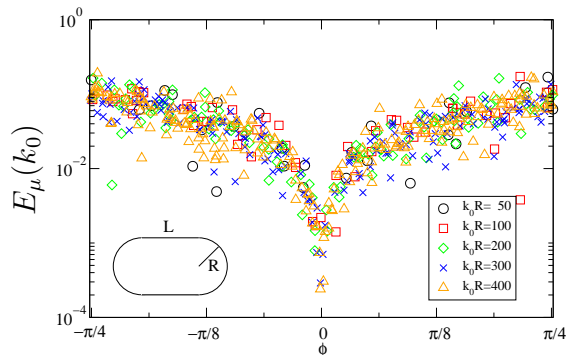}
\caption{Extrapolation error defined by equation~\eqref{eq:error} for eigenfunctions of the stadium billiard of deformation $L/R=2$ (see inset). The horizontal axis represents the initial value of the phase of the eigenvalue $\phi_\mu^{0}$ from which the solution is extrapolated. Data sets for $kR_0=50,100,200,300,400$ are plotted.}
\label{fig:boundaryErr1}
\end{figure}

Building on these observations, we propose the following numerical algorithm to compute both the billiard eigenvalues {\em and} the corresponding eigenfunctions (equation~\eqref{eqlaplaceeig1}). We first determine the unitary eigenvectors, the eigenvalues and their corresponding phase speeds $\{ \lambda_\mu(k_0), u_\mu(k_0), v_\phi^{\mu}(k_0) \}$ at a value $k=k_0$ {\em viz} equation~\eqref{eqeigKsimp1}. This requires two diagonalizations. We then extrapolate the quantization values $k_\mu$ using the approximate constancy of the phase speeds 
\begin{equation}
k_\mu = k_0 + \frac{1}{v_\phi^{\mu}(k_0)} (2\pi - \phi_\mu^{0}),
\end{equation}
where $\phi_\mu^{0} = arg[\lambda_\mu(k_0)]$. The billiard eigenfunctions in the domain are then computed using the approximate $u_\mu(k_0)$ through
\begin{equation}
\psi_\mu(\bm{r}) = \oint_{\partial D} d\bm{r}'(s) G_0(\bm{r},\bm{r}'(s); k_\mu) u_\mu(k_0;s).
\label{eq:wfexpol}
\end{equation}
To assess the accuracy of the solutions we introduce the following quantity, the {\it extrapolation error},
\begin{equation}
E_\mu(k_0) = \frac{|| ( 1 - K(k_\mu) ) u_\mu(k_0) ||_2}{||u_\mu(k_0) ||_2}. \label{eq:error}
\end{equation}
Here, $|| \cdot ||_2$ denotes the 2-norm. In figure~\ref{fig:boundaryErr1}, we plot the resulting error $E_\mu(k_0)$ for extrapolation from various values of initial $k_0$. Instead of $k_0$, we plot the error as a function of $\phi_\mu^{0}$. This provides a measure of the accuracy of the solutions as a function of the interval over which we extrapolate. This in turn determines the fraction of eigenvalues with a given accuracy. Note that a given $\phi_\mu^{0}$ occurs at a different values of $k$ for each $\mu$. 

\begin{figure}[t]
\flushright
\includegraphics[width=13.cm, clip]{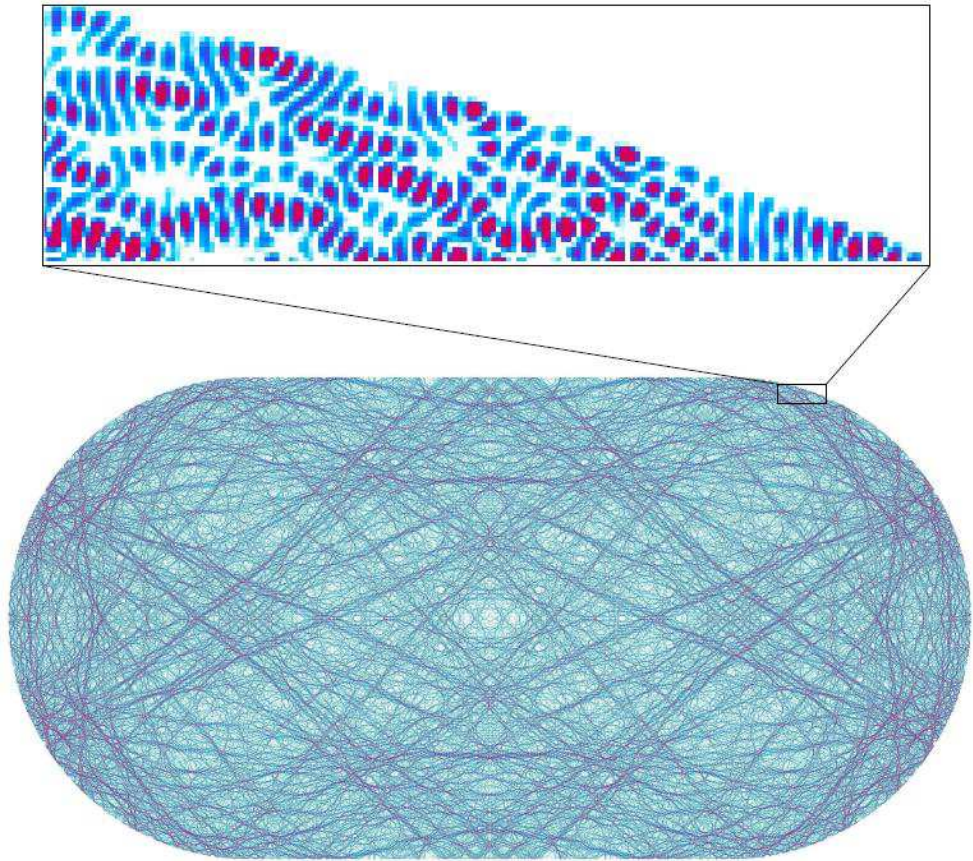}
\caption{A representative plot of a quantized wavefunction for the stadium of deformation $L/R=2$ at $kR=1300.02749$ and an extrapolation error of $0.00296$.}
\label{fig:stadium}
\end{figure}

The data for different $k_0R$ in figure~\ref{fig:boundaryErr1} demonstrates that despite the highly oscillatory nature of the higher lying excited billiard eigenfunctions, the error remains relatively constant as $k$ is increased. A representative highly excited stadium state is plotted in figure~\ref{fig:stadium}.

We should note that the billiard eigenfunctions presented here are domain-normalized. As the normal derivative of the wavefunction $u(s)$ on the boundary contains all information to determine the wavefunction throughout the domain, it's possible to express the normalization condition in terms of $u(s)$ as~\cite{BerryW84} 
\begin{equation}
\oint_{\partial D} ds \, \bm{n}\cdot \bm{r}(s) |u_\mu (k_0;s)|^2 = 2k_0^2,
\end{equation}
which then yields a $\psi_\mu(x)$ which is normalized to unity in $D$.

\begin{table}[t]
\flushright
\caption{Comparison of the EVD and SVD method. The EVD extrapolation is performed at $k_0R=50$. Column 1 contains the extrapolated billiard eigenvalues for a stadium of deformation $L/R=2$. Columns 2-5 contain the eigenvalues obtained by an SVD scan at $25$, $75$, $200$, and $5000$ points in the interval $[40.75,50.25]$. Column 6 contains the extrapolation error for the EVD eigenvalues in Column 1, and the final column contains the relative error of the EVD eigenvalues with respect to SVD5000. In the last row  we have the average relative error for eigenvalues in Columns 1-5 compared to SVD5000. We also quote the computation time on a quad core CPU running at 1.6GHz in the second row. Note that in this table, we show only a fraction of the eigenvalues computed. }\label{tab:EVDsvd}\small
\begin{tabular}{|r|r|r|r|r|r|r|}
\hline
   EVD    &  SVD25    & SVD75     & SVD200   & SVD5000  & $E_\mu$    & Rel.\ error	\\
	10.41s&  35.22s   & 102s      & 272s     & 6728s    &  	         & 		\\
\hline	    										 
49.77668  &  49.7700  &  49.7700  & 49.7725  & 49.77180 &$6.3505E\!-\!3$&$9.8116E\!-\!5$	 \\
49.88509  &  49.8900  &  49.8900  & 49.8850  & 49.88560 &$1.2706E\!-\!2$&$1.0193E\!-\!5$	\\
49.89177  &   ---	  & 	---   & 49.8900  & 49.88900 &$3.7417E\!-\!2$&$5.5620E\!-\!5$	\\
49.94180  &  49.9500  & 	---   & 49.9400  & 49.93930 &$1.6519E\!-\!2$&$5.0195E\!-\!5$	\\
49.94993  &   ---	  &  49.9500  & 49.9500  & 49.94940 &$7.4363E\!-\!3$&$1.0725E\!-\!5$	\\
50.03633  &   ---	  &  50.0430  & 50.0350  & 50.03520 &$1.1042E\!-\!2$&$2.2719E\!-\!5$	\\
50.04363  &  50.0500  & 	---   & 50.0450  & 50.04460 &$5.5522E\!-\!3$&$1.9338E\!-\!5$	\\
50.08058  &   ---	  &  50.0770  & 50.0775  & 50.07810 &$1.0779E\!-\!2$&$4.9640E\!-\!5$	\\
50.08856  &  50.0900  &  50.0900  & 50.0900  & 50.09020 &$1.8785E\!-\!2$&$3.2551E\!-\!5$	\\
50.16631  &  50.1500  &  50.1570  & 50.1575  & 50.15740 &$5.5390E\!-\!2$&$1.7773E\!-\!4$	\\
50.20171  &  50.1900  &  50.1970  & 50.1950  & 50.19410 &$3.6940E\!-\!2$&$1.5175E\!-\!4$	\\
50.23331  &   ---	  &  50.2300  & 50.2325  & 50.23140 &$7.6969E\!-\!2$&$3.8103E\!-\!5$	\\
\hline\hline										
$5.9723E\!-\!5$&$1.7245E\!-\!4 $&$5.7066E\!-\!5 $&$1.1829E\!-\!5$&	0       &            &   		\\
\hline
\end{tabular}
\end{table}

\begin{figure}[hbt]
\flushright
\psfrag{error}{$\!\!\!\!\!\!\!\!\log |1-\lambda|$}
\psfrag{factor}{$\eta$}
\includegraphics[width=9.cm, clip]{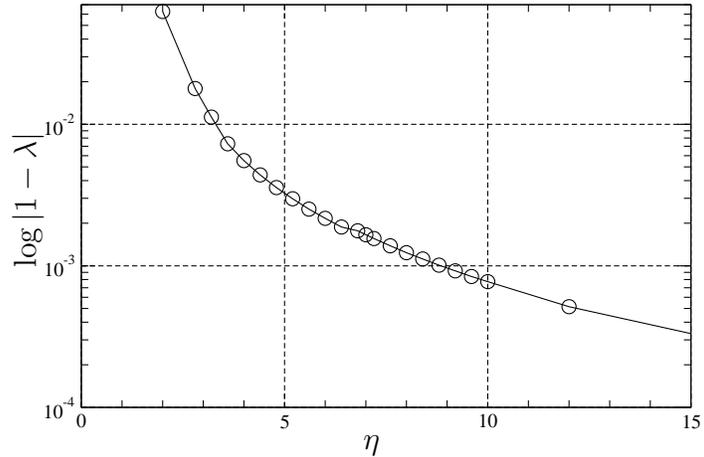}
\caption{Logarithmic plot of the error given by $|1-\lambda_\mu(k_\mu)|$ for a state of the stadium billiard of deformation $L/R=2$ quantized at $kR\approx 20.2965$ as a function of boundary discretization. $\eta$ is defined by $N=\eta kR$}
\label{fig:evdconvergence}
\end{figure}

Next we compare the accuracy of the extrapolation method to that of SVD. In table~\ref{tab:EVDsvd}, we compare the eigenvalues found via EVD extrapolation $k_\mu^{EVD}$ to those found by an SVD scan over an interval of $[40.75,50.25]$. The extrapolation method can attain an accuracy obtained by an SVD scan at about 75 points, providing a factor of roughly 10 in computation speed as seen in table~\ref{tab:EVDsvd}. Important to note is that the simple SVD scan will fail to account for all resonances. Only at a scan over 200 points have all resonances been resolved, which increases the factor to 27.\footnote[7]{As stated before, methods have been proposed to use several of the lowest singular values to resolve nearby resonances~\cite{backer2003}, however we find that such an algorithm still leaves room for ambiguity at large wavevectors compared the the EVD extrapolation method.}  The gain in speed at a fixed accuracy will grow linearly with $kR$ as the number of modes within a given interval of initial phases $\Delta\phi^0$ will increase linearly with $kR$. We would like to emphasize that the accuracy of EVD method and the SVD method is in principle identical (this is clearly seen in comparing the minima attained in figures~\ref{fig:svdevd1}(a) and (b)) and it is the level of accuracy that is desired that will determine the speed enhancement obtained by the EVD method. We have implemented more complex extrapolation methods to provide a desired level of accuracy. 
The ultimate accuracy that can be attained scales exponentially with the number of discretization points on the boundary. This is shown in figure~\ref{fig:evdconvergence}. Finally, in table~\ref{tab:EVDsvdCircle}, we show the accuracy of the EVD method for a case where analytic solutions are available, namely the circular billiard.

\begin{table}
\flushright
\caption{Accuracy of the extrapolation method for the circular billiard. We compare the solutions obtained by the EVD extrapolation method ($k_\mu^{EVD}$) to the solutions obtained analytically ($k_\mu$) by finding the $j^{th}$ zero of the Bessel function $J_m(x)$. We show the relative error (with respect to the analytic solution computed to a precision of $1E\!-\!8$) and the extrapolation error in the fifth and sixth columns, respectively.}\label{tab:EVDsvdCircle}
\begin{tabular}{|r|r|r|r|r|r|r|}
\hline
$m$& $j$&	$k_\mu$	  &$k_\mu^{EVD}$& Rel.\ error  &   $E_\mu$	   \\\hline
6   &1  &  9.93610952 &   9.93723495&$   1.1327E\!-\!4 $ & $  1.8836580E\!-\!3 $   \\
1   &3  & 10.17346813 &  10.17983831&$   6.2616E\!-\!4 $ & $  1.2763550E\!-\!2 $   \\
34  &3  & 49.95933191 &  49.96813288&$   1.7616E\!-\!4 $ & $  1.3162860E\!-\!2 $   \\
16  &9  & 50.04460601 &  50.05588673&$   2.2541E\!-\!4 $ & $  2.1667890E\!-\!2 $   \\
85  &2  & 99.98282066 &  99.98522643&$   2.4062E\!-\!5 $ & $  3.1056810E\!-\!3 $   \\
60  &8  & 99.98510243 &  99.98854423&$   3.4423E\!-\!5 $ & $  6.0242240E\!-\!3 $   \\
24  &21 & 99.99434362 &  99.99440486&$   6.1246E\!-\!7 $ & $  6.0912910E\!-\!4 $   \\
2   &46 &149.99854919 & 149.98312449&$   1.0283E\!-\!4 $ & $  3.0056170E\!-\!2 $   \\	 
68  &19 &150.02814761 & 150.01975466&$   5.5943E\!-\!5 $ & $  1.4191580E\!-\!2 $   \\		  
0   &48 &150.01188245 & 150.02755229&$   1.0446E\!-\!4 $ & $  3.2199340E\!-\!2 $   \\
57  &23 &150.04477281 & 150.04170015&$   2.0478E\!-\!5 $ & $  4.9759550E\!-\!3 $   \\\hline
\end{tabular}						 			
\end{table}

\section{Relation to the SVD method}

\begin{figure}[t]
\flushright
\psfrag{kR}{\raisebox{-1ex}{$kR$}}
\psfrag{absLambda}{min$|\lambda-1|$}
\psfrag{minSigma}{min$\sigma$}
\includegraphics[width=6.5cm]{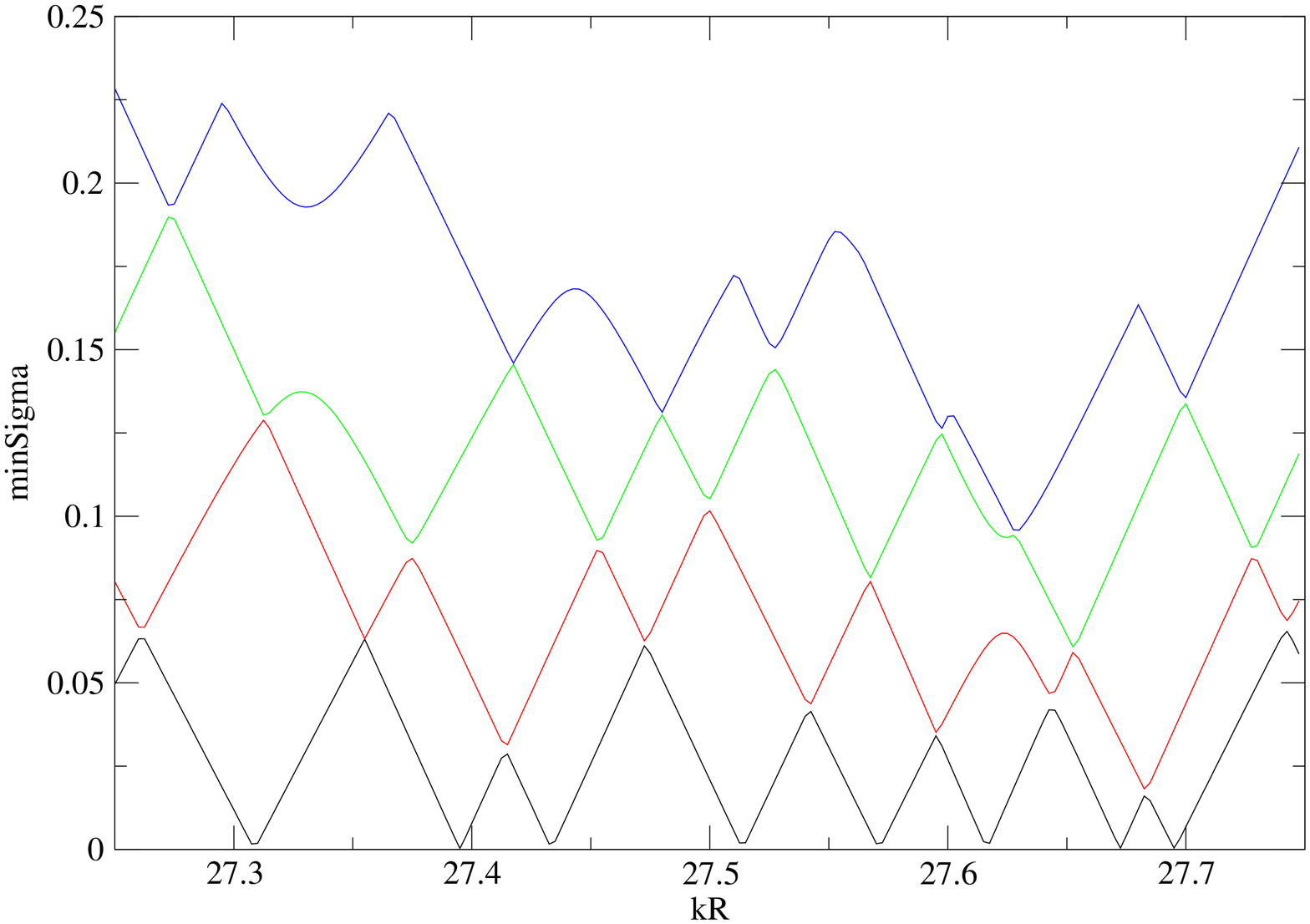}\subfig{-1ex}{5.5cm}{a}
\includegraphics[width=6.5cm]{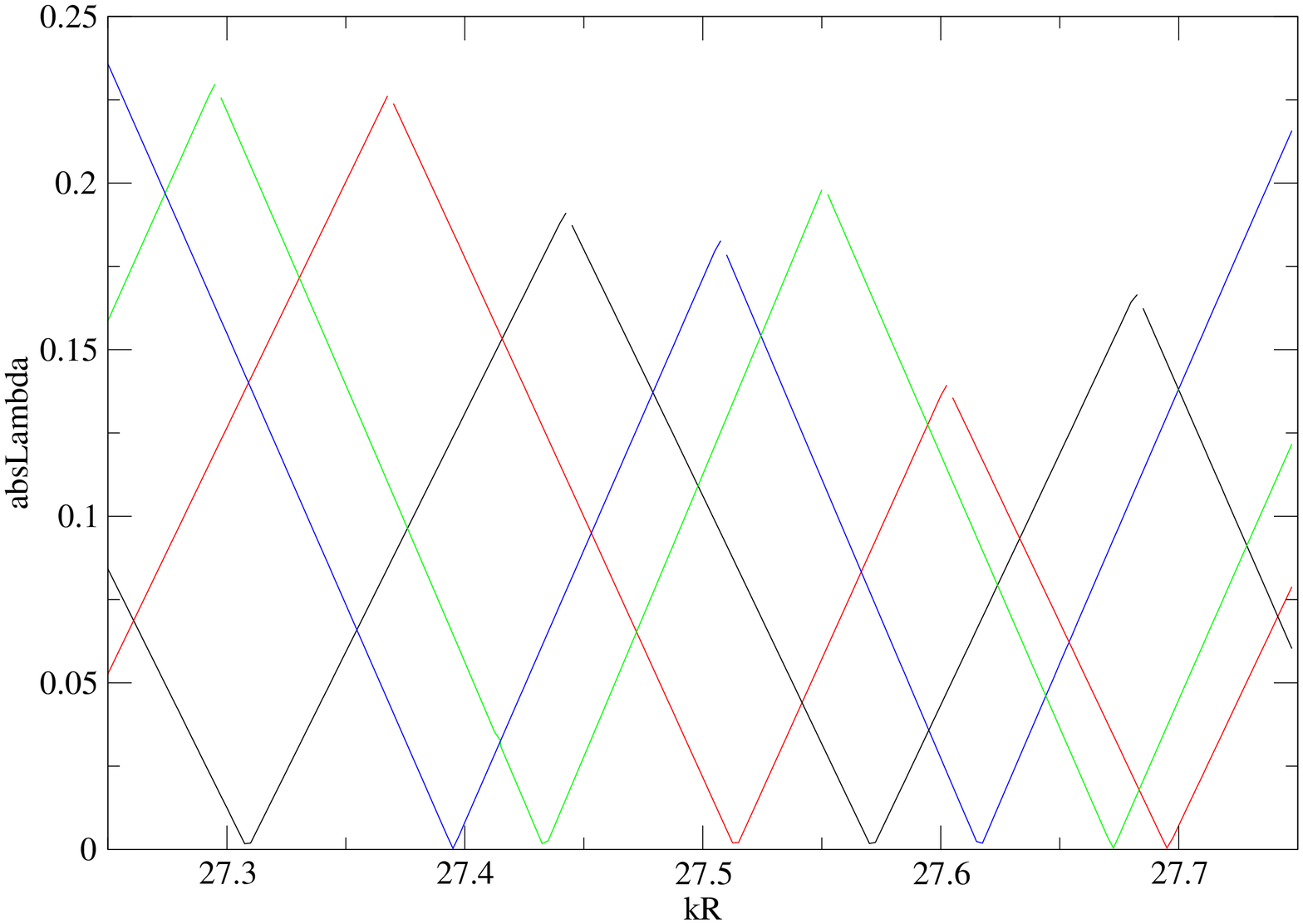}\subfig{-1ex}{5.5cm}{b}
\caption{(a) Comparison of the four lowest singular values $\sigma_\mu(k)$ of $L(k)=1-K(k)$ and (b) the four eigenvalues $\lambda_\mu(k)$ of $K(k)$ with the lowest $|\lambda_\mu(k) -1|$, for a range of $kR=27.25,\ldots,27.75$ in a quadrupole with $\eps=0.3$.}
\label{fig:svdevd1}
\end{figure}

In this section, we would like to clarify the relation between our method and the SVD method~\cite{backer2003, KosztinS97}. In figure~\ref{fig:svdevd1}, we compare the lowest few singular values $\sigma_\mu(k)$ to $| 1 - \lambda_{\mu}(k)|$ which we find by diagonalizing $K(k)$ at an arbitrary $k$ within the spectral range plotted. We find that the plots are almost identical. This should not be surprising, because $L(k)=1-K(k)$ is the matrix whose singular values are computed. A significant point is however that whereas the singular values $\sigma_\mu(k)$ are {\em real} (this is a choice of the numerical SVD routine) and obviously not analytic as a function of $k$, $\lambda_{\mu}(k)$ are complex (and can be shown to be analytic). These points can be put into a more formal setting by following the discussion in~\cite{marks_thesis}. Considering additionally the singular vectors $v$ of $L^{\dagger}(k) = 1-K^{\dagger}(k)$, it can be shown that a generalized {\em complex} singular value can be defined by a proper choice of relative phases of $u_\mu$ and $v_\mu$, which is {\em analytic} as a function of $k$. These analytic complex singular values are exactly  $1 - \lambda_{\mu}(k)$. The real singular value calculated by the numerical SVD routines are $\sigma_\mu(k) = |1 - \lambda_{\mu}(k)|$.

\section{Explicit results in the circular billiard}
In this section we will substantiate the above observations for an analytically solvable problem, namely the Dirichlet problem of a circular billiard. To solve the Laplace eigenvalue problem with Dirichlet boundary conditions for a circular quantum billiard analytically, we can we can write the Green's function, using Bessel addition theorems~\cite{abramovitz}, as
\be
G_0(\bm{r},\bm{r}')= -\frac{i}{4} \beshp{0}{k|\bm{r}-\bm{r}'|} = -\frac{i}{4} \sum_m, \beshp{m}{kr}\besj{m}{kr'} \ex{im(\phi-\phi ')}
\ee
for $r>r'$. Then, assuming $r'$ is on the boundary and $r$ is outside the circular domain, 
\be
\pder{}{G(\bm{r},\bm{r}')}{n'} = \pder{}{G(\bm{r},\bm{r}')}{r'}= -\frac{ik}{4}
\sum_m \mbox{J}_{m}'(kR) \beshp{m}{kr} \ex{im(\phi-\phi')}.
\ee
Let us also Fourier expand the field
\be
u(\phi)=\sum_m q_m \ex{im\phi}.
\ee
Evaluating the integral in~\eqref{eqdirichbim1} we are left with 
a diagonal kernel
\be
K_{mm'} = \left(i\pi kR \, \beshp{m}{kR} \mbox{J}_{m}'(kR)- 1\right) \delta_{mm'}.
\ee
Thus, the singular values can be written as 
\be
\sigma_m(k) = 2 - i\pi kR \, \beshp{m}{kR} \mbox{J}_{m}'(kR) = 0.
\ee	
Using the Bessel identity 
\begin{equation}
\mbox{J}_{m}(x) \mbox{H}^{\pm'}_{m}(x) -\mbox{H}^{\pm}_{m}(x)\mbox{J}_{m}'(x) = \frac{2i}{\pi x},   
\end{equation}
this can be equivalently written as
\begin{equation}
\sigma_m(k) = \mbox{J}_{m}(kR) \mbox{H}^{+'}_{m}(kR).
\end{equation}
Note that the singularity condition yields the secular equation of the {\em internal Dirichlet problem} i.e.\ $\mbox{J}_{m}(kR) = 0$ and that of the {\em external Neumann problem} $\mbox{H}^{+'}_{m}(kR) = 0$ (with Sommerfeld radiation conditions). The latter doesn't have any solutions on the real axis while the former has all its solutions strictly on the real axis. 

\begin{figure}[t]
\centering
\psfrag{Re}{Re$\lambda$}
\psfrag{Im}{Im$\lambda$}
\includegraphics[width=6.5cm]{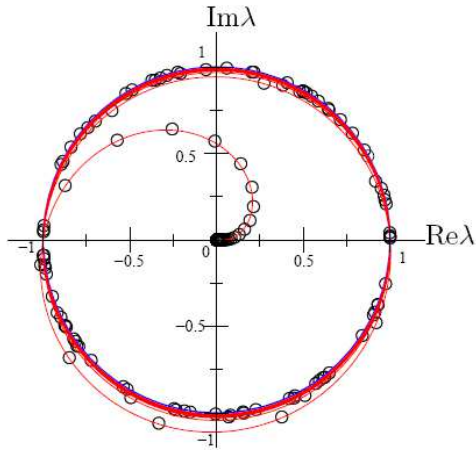}
\caption{The red curve shows the parametric behavior of the complex eigenvalues of the circle $\lambda_m(k)$, via equation~\protect\eqref{eq:analyticEigen} with $m=10$ and $kR=0..100$. In black circles we show the eigenvalues found numerically for $kR=100$ with $N=[10\times kR]$.}
\label{fig:analyticEigen}
\end{figure}

Now let us look at the eigenvalue problem and the extrapolation method. The eigenvalues are parametrically dependent on $k$ and given by 
\begin{equation}
\lambda_m(k) = -1 + i\pi kR \, \beshp{m}{kR} \mbox{J}_{m}'(kR).\label{eq:analyticEigen}
\end{equation}
In figure~\ref{fig:analyticEigen} we show this parametric behavior reproduces the general features observed for smoothly deformed shapes (compare to figure~\ref{figsqevals1}), in particular the transition behavior of eigenvalues from the null-space to the unitary sector.

Using Debye asymptotic expansions of the Bessel Functions, one can show that for $m<kR$ ($m\gg 1$)   
\begin{equation}
\lambda_m(k) \sim \ex{i\Phi},
\end{equation}
where $\Phi = 2m(\tan\beta-\beta) + \pi/2$ and $\cos\beta=m/x$. For $m>kR$, $|\lambda_m| \sim \ex{-2m(\beta - \tan\beta)} \ll 1$. Note that the transition region around $m \sim kR$ (which corresponds to the behavior in the transition region) is not represented uniformly by the above expressions. We thus find that the speed of the unitary eigenvalues (in this case $m<kR$) are asymptotically given by 
\begin{equation}
v_\phi^{m}(k) \sim 2\sin\beta.
\end{equation}
Hence, the change in speed is asymptotically small in $kR$ ($dv_\phi/dk \sim (kR)^{-1}$) as is observed numerically for arbitrary smoothly deformed shapes.

\section{Conclusion}
We have presented an efficient and robust algorithm to calculate eigenvalues of the Laplace operator based on a novel Fredholm formulation. We have shown that approximately of the order of $kR$ eigenvalues can be found with just two diagonalizations and no root search. This overcomes a formidable problem faced by diagonalization algorithms based on SVD for finding large eigenvalues: distinguishing real from false minima in singular values when the density of states $\rho_{Weyl}(k)$ is large.

\ack
This work was partially supported by NSF Grant No. DMR-0408638. HGSL would like to acknowledge the financial support of the Japan Trust during his time at ATR Research Center, Japan, where part of the work was carried out and the financial support of L.\ J.\ Wang. We thank A. Barnett, T.\ Harayama and S.\ Shinohara and A.Douglas Stone for stimulating discussions.
\section*{References}

\begin{thebibliography}{10}

\bibitem{stockmann1992}
J.~{Stein} and H.-J. {St{\"o}ckmann}.
\newblock {Experimental determination of billiard wave functions}.
\newblock {\em Phys.\ Rev.\ Lett.}, 68:2867--2870, May 1992.

\bibitem{blumel1992}
R.~Bl\"umel, I.~H. Davidson, W.~P. Reinhardt, H.~Lin, and M.~Sharnoff.
\newblock Quasilinear ridge structures in water surface waves.
\newblock {\em Phys.\ Rev.\ A}, 45(4):2641--2644, Feb 1992.

\bibitem{soap98}
E.~Arcos, G.~Baez, P.~A. Cuatlayol, M.~L.~H. Prian, R.~A. Mendez-Sanchez, and
  H.~Hernandez-Saldana.
\newblock Vibrating soap films: {An} analog for quantum chaos on billiards.
\newblock {\em Am. J. Phys.}, 66:601, 1998.

\bibitem{science98}
C.~Gmachl, F.~Capasso, E.~E. Narimanov, J.~U. N{\"o}ckel, A.~D. Stone,
  J.~Faist, D.~L. Sivco, and A.~Y. Cho.
\newblock High-power directional emission from microlasers with chaotic
  resonators.
\newblock {\em Science}, 280:1556--1564, 1998.

\bibitem{Doya2001}
Val\'erie Doya, Olivier Legrand, Fabrice Mortessagne, and Christian Miniatura.
\newblock Light scarring in an optical fiber.
\newblock {\em Phys.\ Rev.\ Lett.}, 88(1):014102, Dec 2001.

\bibitem{Reichl2003}
G.~B. {Akguc} and L.~E. {Reichl}.
\newblock Direct scattering processes and signatures of chaos in quantum
  waveguides.
\newblock {\em Phys.\ Rev.\ E}, 67(4):046202, April 2003.

\bibitem{kuhl2005}
U.~Kuhl, H.-J. St\"{o}ckmann, and R~Weaver.
\newblock Classical wave experiments on chaotic scattering.
\newblock {\em J.\ Phys.\ A-Math.\ Gen.}, 38(49):10433--10463, 2005.

\bibitem{wiersig2006}
J.~{Wiersig} and M.~{Hentschel}.
\newblock {Unidirectional light emission from high-{Q} modes in optical
  microcavities}.
\newblock {\em Phys.\ Rev.\ A}, 73(3):031802, March 2006.

\bibitem{stockmann_book}
Hans-J\"urgen St\"ockmann.
\newblock {\em Quantum Chaos: {An Introduction}}.
\newblock Cambridge University Press, Cambridge, United Kingdom, 1999.

\bibitem{haake_book}
F.~Haake.
\newblock {\em Quantum Signatures of Chaos}.
\newblock Springer, Berlin, Germany, 2000.

\bibitem{Y2Kproc_book}
Karl-Fredrik Berggren and Sven {\AA}berg, editors.
\newblock {\em Quantum Chaos Y2K Proceedings of Nobel Symposium 116}.
\newblock Phys.\ Scripta, Stockholm, Sweden, 2000.

\bibitem{Kurland2000}
I.~L. {Kurland}, I.~L. {Aleiner}, and B.~L. {Altshuler}.
\newblock {Mesoscopic magnetization fluctuations for metallic grains close to
  the {Stoner} instability}.
\newblock {\em Phys.\ Rev.\ B}, 62:14886--14897, December 2000.

\bibitem{AgamA01}
O.~Agam and B.~L. Altshuler.
\newblock {''Scars"} in parametrically excited surface waves.
\newblock {\em Physica A}, 302:310--317, 2001.

\bibitem{Stafford2003}
J.~{B{\"u}rki}, R.~E. {Goldstein}, and C.~A. {Stafford}.
\newblock Quantum necking in stressed metallic nanowires.
\newblock {\em Phys.\ Rev.\ Lett.}, 91(25):254501, December 2003.

\bibitem{Zelevinsky2006}
Vladimir Zelevinsky and Alexander Volya.
\newblock Quantum chaos and nuclear physics.
\newblock {\em Phys.\ Scripta}, 2006(T125):147--150, 2006.

\bibitem{harayama2005}
T.~{Harayama}, S.~{Sunada}, and K.~S. {Ikeda}.
\newblock {Theory of two-dimensional microcavity lasers}.
\newblock {\em Phys.\ Rev.\ A}, 72(1):013803, July 2005.

\bibitem{tureci2006}
H.~E. {T{\"u}reci}, A.~D. {Stone}, and L.~{Ge}.
\newblock {Theory of the spatial structure of non-linear lasing modes}.
\newblock {\em Preprint}, cond-mat/0610229, October 2006.

\bibitem{Morton2005}
K.~W. Morton and D.~F. Mayers.
\newblock {\em Numerical Solution of Partial Differential Equations: An
  Introduction}.
\newblock Cambridge University Press, New York, NY, USA, 2005.

\bibitem{alex_thesis}
A.~H. Barnett.
\newblock {\em Dissipation in deforming chaotic billiards}.
\newblock PhD thesis, Harvard University, 2000.

\bibitem{backer2003}
A.~B\"acker.
\newblock Numerical aspects of eigenvalue and eigenfunction computations for
  chaotic quantum systems.
\newblock In S.~Graffi M.~Degli~Esposti, editor, {\em Lecture Notes in
  Physics}, volume 618, pages 91--144. Springer, 2003.

\bibitem{DietzEPSU95}
B.~Dietz, J.~P. Eckmann, C.~A. Pillet, U.~Smilansky, and I.~Ussishkin.
\newblock Inside-outside duality for planar billiards -- a numerical study.
\newblock {\em Phys.\ Rev.\ E}, 51:4222--4231, 1995.

\bibitem{Frischat97}
S.~D. Frischat and E.~Doron.
\newblock Quantum phase-space structures in classically mixed systems: {A}
  scattering approach.
\newblock {\em J.\ Phys.\ A-Math.\ Gen.}, 30:3613--3634, 1997.

\bibitem{Tureci05}
H.~E. T{\"u}reci, H.~G.~L. Schwefel, Ph. Jacquod, and A.~Douglas Stone.
\newblock Modes of wave-chaotic dielectric resonators.
\newblock {\em Progress In Optics}, 47, 2005.

\bibitem{atkinson97numerical}
Kendall~E. Atkinson.
\newblock {\em The Numerical Solution of Integral Equations of the Second
  Kind}.
\newblock Cambridge University Press, Cambridge, MA, 1997.

\bibitem{tureci_thesis}
Hakan~E. T{\"u}reci.
\newblock {\em Wave chaos in dielectric resonators: {A}symptotic and numerical
  approaches}.
\newblock PhD thesis, Yale University, New Haven, USA, 2003.

\bibitem{TasakiHS97}
S.~Tasaki, T.~Harayama, and A.~Shudo.
\newblock Interior {Dirichlet} eigenvalue problem, exterior {Neumann}
  scattering problem, and boundary element method for quantum billiards.
\newblock {\em Phys.\ Rev.\ E}, 56:R13--R16, 1997.

\bibitem{Simonotti2000}
Fernando~P. Simonotti and Marcos Saraceno.
\newblock Fredholm methods for billiard eigenfunctions in the coherent state
  representation.
\newblock {\em Phys.\ Rev.\ E}, 61(6):6527--6537, Jun 2000.

\bibitem{Prange1995}
B.~Georgeot and R.~E. Prange.
\newblock Exact and quasiclassical {Fredholm} solutions of quantum billiards.
\newblock {\em Phys.\ Rev.\ Lett.}, 74(15):2851--2854, Apr 1995.

\bibitem{VerginiS95}
E.~Vergini and M.~Saraceno.
\newblock Calculation by scaling of highly excited-states of billiards.
\newblock {\em Phys.\ Rev.\ E}, 52:2204--2207, 1995.

\bibitem{Barnett2006}
A.~H. {Barnett}.
\newblock {Quasi-orthogonality on the boundary for {Euclidean} {Laplace}
  eigenfunctions}.
\newblock {\em Preprint}, math-ph/0601006, January 2006.

\bibitem{Robnik2007}
G.~Veble, T.~Prosen, and M.~Robnik.
\newblock Expanded boundary integral method and chaotic time-reversal doublets
  in quantum billiards.
\newblock {\em New J.\ Phys.}, 9(1):15, 2007.

\bibitem{Boasman94}
P.~A. Boasman.
\newblock Semiclassical accuracy for billiards.
\newblock {\em Nonlinearity}, 7:485--537, 1994.

\bibitem{KosztinS97}
I.~Kosztin and K.~Schulten.
\newblock Boundary integral method for stationary states of two-dimensional
  quantum systems.
\newblock {\em Int.\ J.\ Mod.\ Phys.\ C}, 8:293--325, 1997.

\bibitem{bogomolny92}
E.~B. Bogomolny.
\newblock Semiclassical quantization of multidimensional systems.
\newblock {\em Nonlinearity}, 5:805--866, 1992.

\bibitem{Doron92}
E.~Doron and U.~Smilansky.
\newblock Semiclassical quantization of chaotic billiards -- a scattering
  theory approach.
\newblock {\em Nonlinearity}, 5:1055--1084, 1992.

\bibitem{Frischat95}
S.~D. Frischat and E.~Doron.
\newblock Semiclassical description of tunneling in mixed systems: case of the
  annular billiard.
\newblock {\em Phys.\ Rev.\ Lett.}, 75:3661, 1995.

\bibitem{BerryW84}
M.~V. Berry and M.~Wilkinson.
\newblock Diabolical points in the spectra of triangles.
\newblock {\em P.\ Roy.\ Soc.\ Lond.\ A Mat.}, 392:15--43, 1984.

\bibitem{marks_thesis}
R.~Bradley Marks.
\newblock {\em Formulation and solution of the open resonator problem}.
\newblock PhD thesis, Yale University, New Haven, USA, 1988.

\bibitem{abramovitz}
M.~Abramovitz and I.~A. Stegun.
\newblock {\em Handbook of mathematical functions}.
\newblock Dover, New York, USA, 1972.

\end{thebibliography}

\end{document}